\documentclass[structabstract]{aa}  
\usepackage{graphicx}
\usepackage{txfonts}
\usepackage{natbib}
\usepackage[breaklinks=false]{hyperref}
\newcommand\cst{{\ensuremath{\mathrm{cst}}}}
\newcommand\dif{{\ensuremath{\mathrm{d}}}}

\newcommand\ds{{\ensuremath{\dif s}}}
\newcommand\dt{{\ensuremath{\dif t}}}
\newcommand\dr{{\ensuremath{\dif r}}}

\newcommand\dP{{\ensuremath{\dif P}}}

\newcommand\dMr{{\ensuremath{\dif M_r}}}
\newcommand\dth{{\ensuremath{\dif\theta}}}
\newcommand\dphi{{\ensuremath{\dif\varphi}}}
\newcommand\dpsi{{\ensuremath{\dif\psi}}}
\newcommand\dxi{{\ensuremath{\dif\xi}}}
\newcommand\Ms{{\ensuremath{\mathrm{M}_{\odot}}}}
\newcommand\Rs{{\ensuremath{\mathrm{R}_{\odot}}}}
\newcommand\kb{\ensuremath{k_{\rm B}}}
\newcommand\mh{\ensuremath{m_{\rm H}}}
\newcommand\asb{\ensuremath{a_{\rm SB}}}
\newcommand\rhc{\ensuremath{\rho_c}}

\newcommand{\Mpy}{\Ms\,{\rm yr}{\ensuremath{^{-1}}}}
\newcommand{\dm}{\ensuremath{\dot M}}
\newcommand{\gva}{{\sc genec}}
\newcommand{\gaz}{{\ensuremath{_{\rm gas}}}}
\newcommand{\rad}{{\ensuremath{_{\rm rad}}}}
\newcommand{\pgaz}{{\ensuremath{P\gaz}}}
\newcommand{\prad}{{\ensuremath{P\rad}}}

\newcommand{\srad}{{\ensuremath{s\rad}}}

\newcommand{\phicore}{{\ensuremath{\varphi_{\rm core}}}}
\newcommand{\phisurf}{{\ensuremath{\varphi_{\rm surf}}}}

\newcommand{\Mcore}{{\ensuremath{M_{\rm core}}}}
\newcommand{\Jcore}{{\ensuremath{J_{\rm core}}}}
\newcommand{\Icore}{{\ensuremath{I_{\rm core}}}}
\newcommand{\Ocore}{{\ensuremath{\Omega_{\rm core}}}}
\newcommand{\Mcrit}{{\ensuremath{M_{\rm crit}}}}
\newcommand{\Mcorecrit}{{\ensuremath{M_{\rm core,crit}}}}

\newcommand{\omgam}{{\ensuremath{\Omega\Gamma}}}
\newcommand{\OmK}{{\ensuremath{\Omega_K}}}
\newcommand{\OmKK}{{\ensuremath{\Omega_K^2}}}
\newcommand{\Omtilde}{{\ensuremath{\tilde{\Omega}}}}
\newcommand{\omtilde}{{\ensuremath{\tilde{\omega}}}}
\newcommand{\gtilde}{{\ensuremath{\tilde{g}}}}
\newcommand{\gbar}{{\ensuremath{\bar{g}}}}

\usepackage{color}

\begin{document}

\title{General-relativistic instability in rapidly accreting supermassive stars: the impact of rotation}

\author{L. Haemmerl\'e}
\authorrunning{Haemmerl\'e}

\institute{D\'epartement d'Astronomie, Universit\'e de Gen\`eve, chemin des Maillettes 51, CH-1290 Versoix, Switzerland}


 
\abstract
{
Supermassive stars (SMSs) collapsing via the general-relativistic (GR) instability are invoked as the possible progenitors of supermassive black holes.
Their mass and angular momentum at the onset of the instability are key in many respects,
in particular regarding the possibility for observational signatures of direct collapse.
Accretion dominates the evolution of SMSs and, like rotation, it has been shown to impact significantly their final properties.
However, the combined effect of accretion and rotation on the stability of these objects is not known.
}
{
Here, we study the stability of rotating, rapidly accreting SMSs against GR and derive the properties of these stars at death.
}
{
On the basis of hylotropic structures, relevant for rapidly accreting SMSs,
we define rotation profiles under the assumption of local angular momentum conservation in radiative regions,
which allows for differential rotation.
We account for rotation in the stability of the structure by adding a Newtonian rotation term in the relativistic equation of stellar pulsation,
which is justified by the slow rotations imposed by the \omgam-limit.
}
{
We find that rotation favours the stability of rapidly accreting SMSs
as soon as the accreted angular momentum represents a fraction $f\gtrsim0.1\%$ of the Keplerian angular momentum.
For $f\sim0.3-0.5\%$ the maximum masses consistent with GR stability are increased by an order of magnitude compared to the non-rotating case.
For $f\sim1\%$, the GR instability cannot be reached if the stellar mass does not exceed $10^7-10^8$ \Ms.
}
{
These results imply that, like in the non-rotating case, the final masses of the progenitors of direct collapse black holes
range in distinct intervals depending on the scenario considered:
$10^5\ \Ms\lesssim M\lesssim10^6$~\Ms\ for primordial atomically cooled haloes;
$10^6\ \Ms\lesssim M\lesssim10^9$~\Ms\ for metal-rich galaxy mergers.
The models suggest that the centrifugal barrier is inefficient to prevent the direct formation of a supermassive black hole at the collapse of a SMS.
Moreover, the conditions of galaxy mergers appear as more favorable than those of atomically cooled haloes
for detectable gravitational wave emission and ultra-long gamma-ray bursts at black hole formation.
}

 
\maketitle
%

\section{Introduction}
\label{sec-in}

The possibility for stars with masses $\gtrsim10^5$ \Ms\ to exist in a stable state of equilibrium has been considered since half a century,
in connection with the physics of quasars
(e.g.~\citealt{hoyle1963a,hoyle1963b,fowler1966,bisnovatyi1967,appenzeller1971a,appenzeller1971b,appenzeller1972a,appenzeller1972b,shapiro1979,fuller1986,baumgarte1999a,baumgarte1999b,begelman2010,hosokawa2013,umeda2016,woods2017,haemmerle2018a,haemmerle2018b,haemmerle2019c};
see \citealt{woods2019} and \citealt{haemmerle2020a} for recent reviews).
One of the particularities of these supermassive stars (SMSs) is to end their life through the general-relativistic (GR) instability \citep{chandrasekhar1964}.
The pressure support of SMSs is dominated by radiation ($\sim99\%$), so that these stars are always close to the Eddington limit.
Since in Newtonian gravity the Eddington limit corresponds to a state of marginal stability,
small GR corrections, of the order of a percent, are sufficient to make the star unstable with respect to radial pulsations.
For Population III (Pop III) SMSs, the subsequent collapse is expected to lead to the direct formation of a supermassive black hole
\citep{fricke1973,shapiro1979,shibata2002,shapiro2002,liu2007,uchida2017,sun2017,sun2018}.

Such direct collapse black holes are invoked to explain the existence of the most massive quasars observed around redshift 7
(e.g.~\citealt{rees1978,rees1984,volonteri2010,volonteri2010b,valiante2017,zhu2020}).
The inferred black hole masses in these objects are in excess of $10^9$ \Ms,
so that accretion must have proceeded at average rates $\gtrsim1-10$ \Mpy\ because of the finite age of the Universe.
The most studied channel of direct collapse is the case of primordial, atomically cooled haloes,
in which mass inflows of 0.1 -- 10 \Mpy\ are expected below a parsec
(e.g.~\citealt{bromm2003b,dijkstra2008,latif2013e,regan2016a,regan2017,chon2018,patrick2020}).
Larger inflows, up to $\sim10^5$ \Mpy, have been found in the simulations of the mergers of massive galaxies at redshifts 8~--~10
\citep{mayer2010,mayer2015,mayer2019}.
Such galaxies are thought to have experiment significant star formation,
so that the progenitors of direct collapse black holes in this channel might be Population I (Pop I) SMSs.
The hydrodynamical simulations of unstable SMSs show that for metallicities as high as solar
the collapse triggers in general a thermonuclear explosion that prevents black hole formation \citep{appenzeller1972a,appenzeller1972b,fuller1986,montero2012}.
Only for masses $\gtrsim10^6$~\Ms\ the deep potential well allows for gravity to bind the collapsing star.
In both cases, direct collapse or supernova explosion, the high energies involved in the death of a SMS
might allow for observational signatures of the existence of these objects.
The simulations of collapsing Pop III SMSs suggest the possibility for gravitational wave emission and ultra-long gamma-ray bursts
\citep{liu2007,shibata2016b,uchida2017,sun2017,sun2018,li2018}.
But the detectability of such signatures depends sensitively on the mass and rotational properties of the SMS at collapse.
In particular, both gravitational waves and gamma-ray bursts require spherical symmetry to be broken, so that rotation is crucial with that respect.

The properties of SMSs at the onset of GR instability have been studied in the case of polytropic structures
(e.g.~\citealt{fowler1966,bisnovatyi1967,baumgarte1999b,shibata2016a,butler2018}).
Stars with masses $\sim10^8-10^9$ \Ms\ are found to be stable, provided rapid enough rotation.
Such structures correspond to fully convective, thermally relaxed stars, assumed to have formed at once ('monolithic' SMSs, \citealt{woods2020}).
They are found to feature universal rotational properties, in particular a universal value for the spin parameter,
which is key regarding black hole formation and gravitational wave emission.
But polytropic models become irrelevant when we account for the formation process of SMSs.
The minimum accretion rates $\sim1$ \Mpy\ that are required to explain the existence of the most massive and distant quasars
are also a necessity for the formation of a SMS, because the H-burning timescale near the Eddington limit is expected to be of the order of the million years,
so that gathering $10^5$~\Ms\ before fuel exhaustion requires rates $\gtrsim0.1$~\Mpy.
Accretion at such rates impacts critically the structure of SMSs, by preventing the thermal relaxation of their envelope
\citep{begelman2010,hosokawa2012a,hosokawa2013,schleicher2013,sakurai2015,haemmerle2018a,haemmerle2019c}.
The high entropy in the outer regions stabilises most of the star against convection,
which develops only in the inner $\sim10\%$ of the mass, triggered by H-burning.
The stability of such structures against GR has been addressed in several works \citep{umeda2016,woods2017,haemmerle2020c,haemmerle2021a}.
SMSs accreting at rates 0.1 -- 10 \Mpy\ are found to collapse at masses $1-4\times10^5$ \Ms,
while masses in excess of $10^6$~\Ms\ would require rates $\gtrsim1000$ \Mpy.
However, these results do not account for the effect of rotation.

Accretion onto a hydrostatic structure requires low enough angular momentum
for the centrifugal force to be cancelled by an excess of gravity over the pressure forces.
The rotation velocity of SMSs is constrained by the \omgam-limit \citep{maeder2000},
which imposes velocities less than $\sim10\%$ of the Keplerian values for masses $\gtrsim10^5$~\Ms,
due to the prominent role of radiation pressure in this mass-range \citep{haemmerle2018b,haemmerle2019a}.
For such slow velocities, rotation hardly impacts the stellar structure, and in particular rotational flattening is negligible.
As we will see, the dynamical effects of rotation are of the same order as those of gas pressure and GR corrections,
so that, as these effects, rotation plays a key role in the stability of the star, but can be neglected regarding its equilibrium properties.

In the present work, we use this advantage to define rotation profiles on the non-rotating structures of rapidly accreting SMSs,
on the basis of the hylotropic models of \cite{begelman2010}.
In the absence of convection, large differential rotation is expected to develop in the interior of SMSs \citep{haemmerle2018b,haemmerle2019a}.
Thus, while solid rotation must be imposed in the convective core,
we will assume that each radiative layer contracts with local angular momentum conservation.
In \cite{haemmerle2020c}, we applied the relativistic pulsation equation of \cite{chandrasekhar1964} to non-rotating hylotropes.
Here, we add to this equation a term accounting for rotation, following the method of \cite{fowler1966}, and apply it to the rotating hylotropes.
The hydrostatic structures and rotation profiles are defined in sections~\ref{sec-hylo} and \ref{sec-omega}, respectively.
The rotation term for the GR instability is introduced in section~\ref{sec-gr} (see also appendix~\ref{app-euler}).
The numerical results are presented in section~\ref{sec-res} and their implications are discussed in section~\ref{sec-dis}.
We summarise the main conclusions in section~\ref{sec-out}.

\section{Method}
\label{sec-meth}

\subsection{Hylotropic structures}
\label{sec-hylo}

Classical spherical stellar structures are determined by the equations of hydrostatic equilibrium and continuity of mass in the following form:
\begin{eqnarray}
{\dP\over\dr}&=&-\rho{GM_r\over r^2}	\label{eq-hydro}\\
{\dMr\over\dr}&=&4\pi r^2\rho			\label{eq-continu}
\end{eqnarray}
where $r$ is the radial distance, $M_r$ the mass enclosed inside $r$, $P$ the pressure, $\rho$ the density of mass and $G$ the gravitational constant.
The structure is fully determined, provided an additional constraint $f(r,M_r,P,\rho)=0$, e.g. a constraint on the entropy profile $s(P,\rho)=s(M_r)$.
In convective regions, where entropy is uniformly distributed, a constraint is given by $s(M_r)=\cst$.
But the entropy gradients that characterise radiative regions prevent in general to close the equation system
without the use of additional differential equations, energy transfer and conservation.

For rapid enough accretion, the entropy profile of SMSs is well described by a hylotropic law \citep{begelman2010,haemmerle2019c,haemmerle2020c}:
\begin{equation}
s=\left\{\begin{array}{cc}	s_c\quad\quad\ \ \quad				&	{\rm if}\ M_r<\Mcore	\\
					s_c\left({M_r\over\Mcore}\right)^{1/2}	&	{\rm if}\ M_r>\Mcore	\end{array}\right.
\label{eq-shylo}\end{equation}
where \Mcore\ is the mass of the isentropic, convective core and $s_c$ the specific entropy in the core.
Equation (\ref{eq-shylo}) provides the required constraint $s(P,\rho)=s(M_r)$ that closes (\ref{eq-hydro}-\ref{eq-continu}),
with a function $s(P,\rho)$ given by the equation of state.

The entropy of SMSs is dominated by radiation,
\begin{equation}
s\simeq\srad={4\over3}{\asb T^3\over\rho},
\label{eq-srad}\end{equation}
where $T$ is the temperature and \asb\ the Stefan-Boltzmann constant.
The contribution from gas reduces to its contribution to pressure:
\begin{equation}
P=\pgaz+\prad={\kb\over\mu\mh}\rho T+{1\over3}\asb T^4,
\label{eq-eos}\end{equation}
where $\mu$ is the mean molecular weight, \mh\ the mass of a proton and \kb\ the Boltzmann constant.
This contribution is expressed by the ratio ($\sim1\%$)
\begin{equation}
\beta:={\pgaz\over P}\simeq{\pgaz\over\prad}={3\over\asb}{\kb\over\mu\mh}{\rho\over T^3}\simeq{\kb\over\mu\mh}{4\over s}.
\label{eq-beta}\end{equation}
Inserting (\ref{eq-shylo}) into (\ref{eq-beta}), we obtain
\begin{equation}
\beta=\left\{\begin{array}{cc}	\beta_c\qquad\qquad				&	{\rm if}\ M_r<\Mcore	\\
						\beta_c\left({M_r\over\Mcore}\right)^{-1/2}	&	{\rm if}\ M_r>\Mcore	\end{array}\right.
\label{eq-bhylo}\end{equation}
where $\beta_c={\kb\over\mu\mh}{4\over s_c}$.
Pressure can be expressed as a function of $\rho$ and $\beta$ instead of $\rho$ and $T$,
using the definition (\ref{eq-beta}) of $\beta$ to eliminate $T$ in (\ref{eq-eos}):
\begin{equation}
P\simeq\left({3\over\asb}\right)^{1/3}\left({\kb\over\mu\mh}\right)^{4/3}{\rho^{4/3}\over\beta^{4/3}}
\label{eq-Pbeta}\end{equation}
Inserting (\ref{eq-bhylo}) into (\ref{eq-Pbeta}), we obtain
\begin{equation}
P=\left\{\begin{array}{cc}	K\rho^{4/3}\qquad\qquad					&	{\rm if}\ M_r<\Mcore	\\
					K\left({M_r\over\Mcore}\right)^{2/3}\rho^{4/3}	&	{\rm if}\ M_r>\Mcore	\end{array}\right.
\label{eq-hylo}\end{equation}
where
\begin{equation}
K=\left({3\over\asb}\right)^{1/3}\left({\kb\over\mu\mh}\right)^{4/3}\beta_c^{-4/3}
\label{eq-K}\end{equation}

The power-law (\ref{eq-hylo}) allows to close the system (\ref{eq-hydro}-\ref{eq-continu}),
and so to solve numerically the stellar structure for any length- and mass-scales,
in a way similar to polytropes \citep{begelman2010}.
To that aim, we define dimensionless functions $\xi$, $\theta$, $\varphi$ and $\psi$:
\begin{equation}
\xi=\alpha r,	\quad
\rho=\rho_c\theta^3,	\quad
M_r={4\over\sqrt{\pi}}\left({K\over G}\right)^{3/2}\varphi,	\quad
P=P_c\psi.
\label{eq-adim}\end{equation}
The length-scale is defined as a function of the central density and pressure
\begin{equation}
\alpha^2={\pi G\rho_c^2\over P_c}={\pi G\rho_c^{2/3}\over K}
\label{eq-alpha}\end{equation}
where we used the fact that $P_c=K\rhc^{4/3}$ (equation~\ref{eq-hylo}).
We notice that the mass-scale is directly given by the constant $K$, i.e. by $\beta_c$ and $\mu$ (equation~\ref{eq-K}).
It can be written alternatively as
\begin{equation}
{4\over\sqrt{\pi}}\left({K\over G}\right)^{3/2}=4\pi\left({K\over\pi G}\right)^{3/2}={4\pi\rhc\over\alpha^3}
\label{eq-mscale}\end{equation}
In these dimensionless quantities, equations (\ref{eq-hydro}-\ref{eq-continu}) and constraint (\ref{eq-hylo}) read
\begin{eqnarray}
{\dpsi\over\dxi}&=&-4{\varphi\theta^3\over\xi^2}	\label{eq-hydroadim}\\
{\dphi\over\dxi}&=&\xi^2\theta^3			\label{eq-continuadim}
\end{eqnarray}
\begin{equation}
\psi=\left\{\begin{array}{cc}	\theta^4	\qquad\qquad					&	{\rm if}\ \varphi<\phicore	\\
						\left({\varphi\over\phicore}\right)^{2/3}\theta^4	&	{\rm if}\ \varphi>\phicore	\end{array}\right.
\label{eq-hyloadim}\end{equation}
We see that, in contrast to polytropes, a free parameter remains in the dimensionless structure itself, \phicore,
the dimensionless mass of the convective core.
Thus, exploring the full parameter space requires to solve a series of structures,
labeled by the various values of \phicore\ in the interval 0.46 -- 2.02.
These two limits correspond respectively to the limit of gravitational binding and to polytropic structures, where $\phisurf=\phicore$ \citep{begelman2010}.
As shown in \cite{haemmerle2020c}, these Newtonian structures allow for precise determination of the onset point of the GR instability
in non-rotating, rapidly accreting SMSs, thanks to the weakness of GR corrections ($\lesssim1\%$).

\subsection{Rotation profiles}
\label{sec-omega}

Because SMSs are close to the Eddington limit, the rotation of their surface is constrained by the \omgam-limit \citep{maeder2000},
which accounts for radiation pressure, instead of the Keplerian limit, based on the simple balance between gravity and the centrifugal force.
The \omgam-limit is extremely restrictive for SMSs, due to their high Eddington factor (0.9 -- 0.99),
and implies surface velocities $\lesssim10\%$ of the Keplerian velocity \citep{haemmerle2018b}:
\begin{equation}
\Omega\lesssim0.1\OmK,
\end{equation}
\begin{equation}
\OmK=\sqrt{GM_r\over r^3}.
\label{eq-kepler}\end{equation}
For such velocities, the centrifugal acceleration represents less than a percent of the gravitational one:
\begin{equation}
{r\Omega^2\over\left({GM_r\over r^2}\right)}={\Omega^2\over\OmKK}\lesssim0.01
\label{eq-centrifuge}\end{equation}
Interestingly, this dynamical contribution is of the same order as GR corrections and departures from the Eddington limit.
Thus, like these effects, rotation shall play a significant role on the stability of SMSs,
but can be neglected in the definition of the hydrostatic structures.

Here, we use this advantage to define rotation profiles on the non-rotating hylotropes of section~\ref{sec-hylo}.
As a result of the short evolutionary timescales, angular momentum is expected to be locally conserved in the radiative envelope of SMSs
\citep{haemmerle2018b,haemmerle2019a}.
Thus, we will assume that the specific angular momentum of each radiative layer is given by that at accretion.
Then, as the successive layers join the convective core, they advect this angular momentum,
which is redistributed inside the core according to the constraint of solid rotation.
Notice that we neglect the effect of the outer convective zone found in models of rapidly accreting SMSs \citep{hosokawa2013,haemmerle2018a}.
This envelope plays negligible role in the GR instability, as a result of its small mass and low density \citep{haemmerle2020c}.
It plays a critical role in the rotation of the surface, because its small mass covers large radii,
over which angular momentum is instantaneously redistributed \citep{haemmerle2018b,haemmerle2019a}.
However, since it represents a percent of the total stellar mass,
this redistribution impacts only locally the angular momentum profile once the layers contract in the radiative regions.
Indeed, at each time, the specific angular momentum $j_-$ of the mass that leaves the envelope at its bottom $r_-$
is given by the angular velocity $\Omega_{\rm env}$ of the envelope in solid rotation:
\begin{eqnarray}
j_-={2\over3}r_-^2\Omega_{\rm env}={2\over3}r_-^2{J_{\rm env}\over I_{\rm env}}={{2\over3}r_-^2M_{\rm env}\over I_{\rm env}}\langle j_{\rm accr}\rangle
\end{eqnarray}
where $M_{\rm env}$, $J_{\rm env}$ and $I_{\rm env}$ are the mass, the angular momentum and the moment of inertia of the envelope, respectively,
and $\langle j_{\rm accr}\rangle=J_{\rm env}/M_{\rm env}$ is the average specific angular momentum advected at accretion by the layers of the envelope,
that provides its angular momentum content.
The large density contrasts in the envelope imply that most of its mass is located near $r_-$,
so that $I_{\rm env}\simeq{2\over3}r_-^2M_{\rm env}$, and $j_-\simeq\langle j_{\rm accr}\rangle$.
Since the layers of the envelope, over which the average $\langle\ \rangle$ is done, represent only a percent of the total mass,
we can assume $j_{\rm accr}\simeq\cst=\langle j_{\rm accr}\rangle$ for all these layers,
so that the angular momentum $j_-$ advected by a layer in the radiative region is always close to that it advected at accretion: $j_-\simeq j_{\rm accr}$.
In other words, regarding the inner rotation profiles, the convective envelope can be considered as a unique layer located in $r_-$.
Only when we consider the surface velocity the depth of the envelope becomes important.

The accreted angular momentum is chosen as a fraction $f$ of the Keplerian momentum at the equator:
\begin{equation}
j_{\rm accr}=f\sqrt{GMR_{\rm accr}(M)}
\label{eq-jac}\end{equation}
where $R_{\rm accr}(M)$ is the accretion radius, i.e. the radius of the accretion shock when the star has a mass $M$.
We use the mass-radius relation of rapidly accreting SMSs \citep{hosokawa2012a,hosokawa2013,haemmerle2018a,haemmerle2019c}:
\begin{eqnarray}
R_{\rm accr}(M)=260\,\Rs\left({M\over\Ms}\right)^{1/2}.
\label{eq-rac}\end{eqnarray}
This power-law, that reflects the evolution along the Eddington and Hayashi limits, reproduces well the models for masses $\lesssim10^5$~\Ms.
Possible departures from this relation at larger masses are discussed in section~\ref{sec-rotation}.
Notice that the dependence $R\propto M^{1/2}$ implies a constant surface gravity:
\begin{eqnarray}
\gbar={GM\over R_{\rm accr}^2}={1\over260^2}{G\Ms\over\rm R_\odot^2}\simeq0.4\rm\ cm\ s^{-2}
\label{eq-gbar}\end{eqnarray}
The initial angular momentum of each layer $M_r$, conserved until it joins the core,
is given by the angular momentum advected at the accretion of the layer (equation~\ref{eq-jac}), when the mass of the star was $M=M_r$:
\begin{equation}
j(M_r)=f\sqrt{GM_rR_{\rm accr}(M_r)}
\label{eq-jfmr}\end{equation}
With the mass-radius relation (\ref{eq-rac}), it is given uniquely by the mass-coordinate $M_r$ of the layer and the fraction $f$:
\begin{equation}
j(M_r)=f\,{(GM_r)^{3/4}\over\gbar^{1/4}}
\label{eq-jmr}\end{equation}
The angular velocity in the radiative envelope is then given by
\begin{eqnarray}
\Omega(M_r)&=&{3\over2}{j(M_r)\over r^2}={3\over2}f\OmK\left({g\over\gbar}\right)^{1/4}
\label{eq-omrad}\end{eqnarray}
where we used the Keplerian velocity (\ref{eq-kepler}) and the local gravitational acceleration $g=GM_r/r^2$.
We see that the ratio of $\Omega$ to its Keplerian value is given uniquely by $f$ and the local gravity.
The total angular momentum of the core is the integral of the angular momentum advected by the layers $M_r<\Mcore$:
\begin{eqnarray}
\Jcore&=&\int_0^\Mcore j(M_r)\dMr={4\over7}\,f{G^{3/4}M_{\rm core}^{7/4}\over\gbar^{1/4}}
\label{eq-jcore}\end{eqnarray}
From the assumption of solid rotation, the angular velocity of the core is
\begin{eqnarray}
\Ocore&=&{\Jcore\over\Icore}
\label{eq-omcore}\end{eqnarray}
where \Icore, the moment of inertia of the core, is given by the non-rotating structure.

In order to apply these rotation profiles to the dimensionless hylotropic structures of section~\ref{sec-hylo},
we rewrite them with the use of the dimensionless functions (\ref{eq-adim}).
The gravitational acceleration reads
\begin{eqnarray}
g&=&{GM_r\over r^2}={4\pi G\rhc\over\alpha}{\varphi\over\xi^2}=\sqrt{16\pi GP_c}\,{\varphi\over\xi^2}
\end{eqnarray}
It is fully determined by the dimensionless structure $(\xi,\varphi)$ and the central pressure.
At first order, the central pressure is given by radiation, and thus depends on the central temperature only.
The rotation profile in the radiative envelope is given by (equation~\ref{eq-omrad})
\begin{eqnarray}
{\Omega\over\OmK}&=&{3\over2}f\left({\sqrt{16\pi GP_c}\over\gbar}\right)^{1/4}{\varphi^{1/4}\over\xi^{1/2}}
\end{eqnarray}
From equation~(\ref{eq-jcore}), the angular momentum of the core is
\begin{eqnarray}
\Jcore&=&{4\over7}f\left({4\pi\rhc\over\alpha^3}\right)^{7/4}{G^{3/4}\varphi_{\rm core}^{7/4}\over\gbar^{1/4}}
\end{eqnarray}
By definition, the moment of inertia is given by
\begin{eqnarray}
\Icore&=&{2\over3}\int_0^\Mcore r^2\dMr={2\over3}{4\pi\rhc\over\alpha^5}\int_0^\phicore\xi^2\dphi
\end{eqnarray}
Equation~(\ref{eq-omcore}) gives
\begin{eqnarray}
{\Ocore\over\sqrt{4\pi G\rhc}}&=&{6\over7}f\left({\sqrt{16\pi GP_c}\over\gbar}\right)^{1/4}
{\varphi_{\rm core}^{7/4}\over\int_0^\phicore\xi^2\dphi}
\end{eqnarray}
The ratio to the Keplerian velocity (equation~\ref{eq-kepler})
\begin{eqnarray}
\OmK&=&\sqrt{GM_r\over r^3}=\sqrt{4\pi G\rhc}\ {\varphi^{1/2}\over\xi^{3/2}}
\label{eq-kepleradim}\end{eqnarray}
is
\begin{eqnarray}
{\Ocore\over\OmK}
&=&{6\over7}f\left({\sqrt{16\pi GP_c}\over\gbar}\right)^{1/4}{\varphi_{\rm core}^{7/4}\over\int_0^\phicore\xi^2\dphi}{\xi^{3/2}\over\varphi^{1/2}}
\end{eqnarray}
Thus, if we define
\begin{eqnarray}
\gtilde:={\sqrt{16\pi GP_c}\over\gbar}
\label{eq-gtilde}\end{eqnarray}
\begin{eqnarray}
\omtilde:=\left\{\begin{array}{cc}
{\xi^{3/2}\over\varphi^{1/2}}\cdot{{4\over7}\varphi_{\rm core}^{7/4}\over\int_0^\phicore\xi'^2\dif\varphi'}	\qquad			&\rm if\ \varphi<\phicore	\\
{\varphi^{1/4}\over\xi^{1/2}}														\qquad\qquad\qquad	&\rm if\ \varphi>\phicore
\end{array}\right.
\label{eq-omtilde}\end{eqnarray}
we can write the full rotation profile, core+envelope, in the following way:
\begin{eqnarray}
{\Omega\over\OmK}&=&{3\over2}f\gtilde^{1/4}\omtilde
\label{eq-omomk}\end{eqnarray}
The gravity-ratio \gtilde\ is given uniquely by the central pressure, i.e. by the central temperature,
while the dimensionless profile \omtilde\ is given uniquely by the dimensionless hylotropic structure.
Thus, for a given hylotrope, the profile of $\Omega/\OmK$ is given uniquely by the fraction $f$ and the central temperature.
We notice that for the central temperatures of SMSs ($10^7-10^8$ K), the gravity ratio \gtilde\ takes values of $\sim10^6$,
which reflects the strong contraction of the central layers of SMSs from their large accretion radii~(\ref{eq-rac}).
Equation~(\ref{eq-omomk}) suggests that typical values $f\sim0.1-1\%$ are required for the typical velocities $\sim0.1\OmK$ imposed by the \omgam-limit.
The $\Omega$-profile requires in addition the scaling factor for \OmK, given by central density (equation~\ref{eq-kepleradim}):
\begin{eqnarray}
{\Omega\over\sqrt{4\pi G\rhc}}&=&{3\over2}f\gtilde^{1/4}\Omtilde
\label{eq-om}\end{eqnarray}
where
\begin{eqnarray}
\Omtilde:={\varphi^{1/2}\over\xi^{3/2}}\omtilde=\left\{\begin{array}{cc}
{{4\over7}\varphi_{\rm core}^{7/4}\over\int_0^\phicore\xi'^2\dif\varphi'}	\quad	&\rm if\ \varphi<\phicore	\\
{\varphi^{3/4}\over\xi^2}										\ \qquad	&\rm if\ \varphi>\phicore
\end{array}\right.
\label{eq-Omtilde}\end{eqnarray}
From these rotation profiles, we can derive the profiles of the spin parameter:
\begin{eqnarray}
a_r&:=&{cJ_r\over GM_r^2}={c\over GM_r^2}\int_0^{M_r}{2\over3}\Omega'r'^2\dif M_r'	\label{eq-spindef}\\
&=&f\,{c\over\gbar^{1/4}}\left(4\pi G\left({K\over\pi G}\right)^{3/2}\right)^{-1/4}{\int_0^\varphi\Omtilde'\xi'^2\dif\varphi'\over\varphi^2}
\label{eq-spin}\end{eqnarray}
We see that the scaling factor of the spin parameter is given by the mass-scale and the fraction $f$.
Notice that, in the radiative envelope ($M_r>\Mcore$), the spin parameter can be obtained directly by integration of equation~(\ref{eq-jmr}):
\begin{equation}
a_r=f\,{c\over\gbar^{1/4}}{4\over7}\left(GM_r\right)^{-1/4}	\qquad\rm (envelope)
\label{eq-spinaccr}\end{equation}
Since convection redistributes the angular momentum only in the core, the amount $J_r$ included in the mass $M_r$ remains unchanged,
and is set directly by the accretion law~(\ref{eq-jac}), independently of the hylotropic structure.
The choice of the structure impacts the spin parameter only in the core.
For $M_r<\Mcore$, equation~(\ref{eq-spin}) gives
\begin{eqnarray}
a_r&=&f\,{c\over\gbar^{1/4}}{4\over7}\left(GM_r\right)^{-1/4}\cdot\left({M_r\over\Mcore}\right)^{-7/4}{I_r\over\Icore}	\qquad\rm (core)
\label{eq-acore}\end{eqnarray}
where $I_r:={2\over3}\int_0^{M_r}r'^2\dif M_r'$ is the moment of inertia of the mass $M_r$.
The spin parameter is changed by a factor that depends on the two dimensionless ratios $M_r/\Mcore$ and $I_r/\Icore$,
uniquely given by the dimensionless structure.
Thus, the full profile of the spin parameter, in the core and the envelope, is given by the dimensionless structure, the fraction $f$ and the mass-scale.

Finally, the rotational kinetic energy is given by (equation~\ref{eq-Eom})
\begin{eqnarray}
E_\Omega={1\over3}\int{\Omega^2\over\OmKK}{GM_r\over r}\dMr
\label{eq-Eom2}\end{eqnarray}
Its ratio to the absolute value of the gravitational energy
\begin{eqnarray}
\vert W\vert=\int{GM_r\over r}\dMr
\label{eq-Egrav}\end{eqnarray}
corresponds to the average of ${1\over3}\Omega^2/\OmKK$ weighted on the gravitational energy per mass-unit.
With the rotation profile~(\ref{eq-omomk}), we obtain
\begin{eqnarray}
{E_\Omega\over\vert W\vert}={3\over4}f^2\gtilde^{1/2}{\int\omtilde^2{\varphi\over\xi}\dphi\over\int{\varphi\over\xi}\dphi}
={3\over4}f^2\gtilde^{1/2}{\int\omtilde^2\varphi\theta^3\xi\dxi\over\int\varphi\theta^3\xi\dxi}
\label{eq-TWadim}\end{eqnarray}
where we used (\ref{eq-continuadim}) to substitute \dxi\ to \dphi\ in the integrals.
Since $E_\Omega/\vert W\vert$ is an average of $\Omega^2/\OmKK$, these two quantities share the same scaling factor.
As a consequence, the profile of $\Omega/\OmK$ (equation~\ref{eq-omomk}) is fully determined by $E_\Omega/\vert W\vert$ and the dimensionless structure:
\begin{eqnarray}
{\Omega^2\over\OmKK}&=&{3E_\Omega\over\vert W\vert}\,{\omtilde^2\int\varphi\theta^3\xi\dxi\over\int\omtilde^2\varphi\theta^3\xi\dxi}
\label{eq-omomkk}\end{eqnarray}
Equation~(\ref{eq-omomkk}) suggests $E_\Omega/\vert W\vert\sim0.01$ for the velocities $\sim0.1\OmK$ of the \omgam-limit.

\subsection{Impact of rotation on the GR instability}
\label{sec-gr}

The relativistic equation for the frequency $\omega$ of adiabatic pulsations has been derived by \cite{chandrasekhar1964}
from Einstein's equations in full generality, with the only assumption of spherical symmetry.
In \cite{haemmerle2020c,haemmerle2021a}, we have shown that this equation allows to determine with a high precision
the stability of any spherical (i.e. non-rotating) hydrostatic structure with respect to GR, if used in the following form:
\begin{eqnarray}
{\omega^2\over c^2}I_0=\sum_{i=1}^4I_i
\label{eq-chandra}\end{eqnarray}
with
\begin{eqnarray}
I_0&=&\int_0^Re^{a+3b}(P+\rho c^2)r^4\dr					\label{eq-I0}\\
I_1&=&9\int_0^Re^{3a+b}\left(\Gamma_1-{4\over3}\right)Pr^2\dr	\label{eq-I1}\\
I_2&=&-12\int_0^Re^{3a+b}\left(a'+{b'\over3}\right)Pr^3\dr			\label{eq-I2}\\
I_3&=&{8\pi G\over c^4}\int_0^Re^{3(a+b)}P(P+\rho c^2)r^4\dr		\label{eq-I3}\\
I_4&=&-\int_0^Re^{3a+b}{P'^2\over P+\rho c^2}r^4\dr			\label{eq-I4}
\end{eqnarray}
where $'$ indicates the derivatives with respect to $r$ ($r=R$ at the surface),
$\Gamma_1$ is the first adiabatic exponent,
and $a$ and $b$ are the coefficients of the metric $\ds^2=-e^{2a}(c\dt)^2+e^{2b}\dr^2+r^2\dif\Omega^2$, given by:
\begin{eqnarray}
a'&=&{GM_r\over r^2c^2}{1+{4\pi r^3\over M_rc^2}P\over1-{2GM_r\over rc^2}}	,\qquad
e^{2a(R)}=1-{2GM_R\over Rc^2}	\label{eq-a}\\
e^{-2b}&=&1-{2GM_r\over rc^2}	\label{eq-b}
\end{eqnarray}

Thanks to the weakness of GR corrections in SMSs, we can restrict our analysis to the first order post-Newtonian component of equation (\ref{eq-chandra}).
Moreover, due to the proximity with the Eddington limit, equation (\ref{eq-chandra}) can be linearised in $\beta$,
using the following expression for $\Gamma_1$ \citep{haemmerle2020c}:
\begin{equation}
\Gamma_1={4\over3}+{\beta\over6}+\mathcal{O}(\beta^2)
\label{eq-gamma1}\end{equation}
We have shown in \cite{haemmerle2020c} that the first order post-Newtonian and sub-Eddington components
of integrals (\ref{eq-I1}-\ref{eq-I2}-\ref{eq-I3}-\ref{eq-I4}) can be written in the following form
with the help of the dimensionless functions defined in (\ref{eq-adim}):
\begin{eqnarray}
I_1&=&{3\over2}\beta_c{P_c\over\alpha^3}\int{\beta\over\beta_c}\psi\xi^2\dxi				\label{eq-I1adim}\\
I_2&=&-16\sigma{P_c\over\alpha^3}\int\left(\xi\theta^3+{2\varphi\over\xi^2}\right)\psi\xi^3\dxi	\label{eq-I2adim}\\
I_3&=&8\sigma{P_c\over\alpha^3}\int\psi\theta^3\xi^4\dxi								\label{eq-I3adim}\\
I_4&=&-16\sigma{P_c\over\alpha^3}\int\varphi^2\theta^3\dxi							\label{eq-I4adim}
\end{eqnarray}
The dimensionless parameter $\sigma$, defined by \citep{tooper1964a}
\begin{equation}
\sigma:={P_c\over\rhc c^2}={K\rho_c^{1/3}\over c^2}={\pi G\rhc\over\alpha^2c^2},
\label{eq-sigma}\end{equation}
represents the departures from the Newtonian limit, while $\beta_c$ represents the departures from the Eddington limit.
Integrals (\ref{eq-I1adim}-\ref{eq-I2adim}-\ref{eq-I3adim}-\ref{eq-I4adim}) have been obtained by linearisation with respect to these two parameters,
i.e. neglecting terms in $\mathcal{O}(\sigma^2)$, $\mathcal{O}(\beta_c^2)$ or $\mathcal{O}(\sigma\beta_c)$.
Notice that the compactness is related to $\sigma$ by
\begin{equation}
{2GM_r\over rc^2}=8\sigma{\varphi\over\xi}
\label{eq-compact}\end{equation}
Since all integrals (\ref{eq-I1adim}-\ref{eq-I2adim}-\ref{eq-I3adim}-\ref{eq-I4adim}) have a global $\sigma$ or $\beta_c$ factor in front,
their content can be evaluated with Newtonian equations.
In particular, equations (\ref{eq-hydroadim}-\ref{eq-continuadim}) and an integration by parts allow to write
\begin{eqnarray}
\int\psi\theta^3\xi^4\dxi	=\int\psi\xi^2{\dphi\over\dxi}\dxi
					=-\int\varphi{\dif\over\dxi}\left(\xi^2\psi\right)\dxi	\\
					=-2\int\psi\varphi\xi\dxi+4\int\varphi^2\theta^3\dxi,
\end{eqnarray}
so that
\begin{equation}
I_2=-64\sigma{P_c\over\alpha^3}\int\varphi^2\theta^3\dxi=4I_4
\end{equation}
Thus, in the post-Newtonian and sub-Eddington limit, the sum (\ref{eq-chandra}) of the integrals reduces to
\begin{eqnarray}
{2\over3}{\alpha^3\over P_c}{\omega^2\over c^2}I_0=&&\beta_c\int{\beta\over\beta_c}\psi\xi^2\dxi	\nonumber\\
&&-{32\over3}\sigma\left(\int\psi\varphi\xi\dxi+3\int\varphi^2\theta^3\dxi\right)
\label{eq-postnewtonadim}\end{eqnarray}
This sum can be re-expressed in terms of physical quantities, using equations (\ref{eq-adim}-\ref{eq-alpha}) and (\ref{eq-sigma}):
\begin{eqnarray}
{2\over3}{\omega^2\over c^2}I_0
&=&\int\beta Pr^2\dr-{8\over3}{G\over c^2}\int PM_r r\dr-{2G^2\over c^2}\int M_r^2\rho\dr	\\
&=&\int\beta Pr^2\dr-\int\left({2GM_r\over rc^2}+{8\over3}{P\over\rho c^2}\right){GM_r\over r}\rho r^2\dr
\label{eq-postnewton1}\end{eqnarray}
Using the volume $\dif V=4\pi r^2\dr$ and $\dMr$ from (\ref{eq-continu}), the equation becomes
\begin{eqnarray}
\omega^2I&=&\int\beta P\dif V-\int\left({2GM_r\over rc^2}+{8\over3}{P\over\rho c^2}\right){GM_r\over r}\dMr
\label{eq-postnewton2}\end{eqnarray}
where $I={8\over3}{\pi\over c^2}I_0$ reduces to the moment of inertia in the Newtonian limit.

Equation (\ref{eq-postnewton2}) contains all the terms that are required to capture the GR instability on any hydrostatic structure
in the post-Newtonian + sub-Eddington limit, in the case of pure spherical symmetry, i.e. without rotation.
The condition for the instability is an imaginary frequency $\omega^2<0$.
Since $I$ is always positive, it happens when the negative term in the right-hand side, that contains the GR corrections in $c^{-2}$,
exceeds in absolute value the positive term, that contains $\beta$, the departures from the Eddington limit.
Notice that this positive term is purely Newtonian, since crossed GR and $\beta$ corrections would give only second-order terms.
Since $\beta$ is the ratio of gas to total pressure, this integral corresponds actually to the integrated gas pressure,
which is $2/3$ of the total internal energy contained in the gas.
On the other hand, the negative integral, that scales with the GR corrections, scales also with the total gravitational energy of the star,
since the integrand scales with the gravitational energy per mass unit $-GM_r/r$.

Since both $\beta$ and the GR corrections have typical values of 0.01 in SMSs,
the positive and negative integrals in equation (\ref{eq-postnewton2}) represent about a percent of the total internal and gravitational energies, respectively.
As we saw in section~\ref{sec-omega}, the dynamical contribution by rotation is of the same order.
Thus, like the effect of gas pressure, the effect of rotation on the stability of the star can be reduced to its Newtonian component,
since relativistic rotation terms would be of second order.
This method has been used by \cite{fowler1966}, \cite{bisnovatyi1967} and \cite{baumgarte1999b} for polytropic models.
For slow rotation near the Eddington limit, a Newtonian pulsation analysis leads to (see appendix~\ref{app-euler}, equation~\ref{eq-puls1})
\begin{eqnarray}
\omega^2I&=&\int\beta P\dif V+{4\over9}\int{\Omega^2\over\OmKK}{GM_r\over r}\dMr
\label{eq-rotation}\end{eqnarray}
The $\Omega^2$-term in equation (\ref{eq-rotation}) is the rotational energy (\ref{eq-Eom2}) multiplied by a factor 4/3.
Since this term is positive, we see that rotation acts as a stabilising agent, like gas pressure, against the destabilising GR effects.
It is expressed by adding this term in the post-Newtonian equation (\ref{eq-postnewton2}):
\begin{eqnarray}
\omega^2I
&=&\int\beta P\dif V
-\int\left({2GM_r\over rc^2}+{8\over3}{P\over\rho c^2}-{4\over9}{\Omega^2\over\OmKK}\right){GM_r\over r}\dMr
\label{eq-postnewtonomega}\end{eqnarray}
In the sum of the $I_i$ that appears in equation (\ref{eq-chandra}), this term translates into a fifth integral:
\begin{equation}
I_5:={2\over3}\int\Omega^2\rho r^4\dr={8\over3}{P_c\over\alpha^3}\int{\Omega^2\over\OmKK}\varphi\theta^3\xi\dxi
\label{eq-I5}\end{equation}
With the rotation profiles defined in section~\ref{sec-omega} (equation~\ref{eq-omomk}), it gives
\begin{equation}
I_5=6f^2\gtilde^{1/2}{P_c\over\alpha^3}\int\omtilde^2\varphi\theta^3\xi\dxi
\label{eq-I5om}\end{equation}
Adding this term to equation~(\ref{eq-postnewtonadim}), we obtain
\begin{eqnarray}
{2\over3}{\alpha^3\over P_c}{\omega^2\over c^2}I_0
&=&\beta_c\int{\beta\over\beta_c}\psi\xi^2\dxi
+4f^2\gtilde^{1/2}\int\omtilde^2\varphi\theta^3\xi\dxi					\nonumber\\&&\qquad\qquad
-{32\over3}\sigma\left(\int\psi\varphi\xi\dxi+3\int\varphi^2\theta^3\dxi\right)
\label{eq-gr1}\end{eqnarray}
We see that the rotation term scales with the independent fraction $f^2$, and with the square root of the gravity ratio \gtilde.
From its definition~(\ref{eq-gtilde}), $\gtilde^{1/2}$ scales with $P_c^{1/4}$, i.e. with the central temperature:
\begin{eqnarray}
\gtilde^{1/2}(T_c)=\left({\sqrt{16\pi G}\over\gbar}\right)^{1/2}\left({\asb\over3}\right)^{1/4}T_c
\label{eq-gtildeTc}\end{eqnarray}
The product of the other two scaling factors, $\beta_c$ and $\sigma$, is also given by the central temperature and the chemical composition:
\begin{equation}
\beta_c\sigma={\kb T_c\over\mu\mh c^2}
\label{eq-betasigma}\end{equation}
by definitions (\ref{eq-beta}-\ref{eq-sigma}).
Thus, the limit of stability, at which the right-hand side of equation~(\ref{eq-gr1}) vanishes,
is given by a second-order polynomial equation in $\beta_c$:
\begin{eqnarray}
\beta_{c,\rm crit}^2
+4f^2\gtilde^{1/2}(T_c)\,\tilde I_1\cdot\beta_{c,\rm crit}
-{\kb T_c\over\mu\mh c^2}\,\tilde I_0
=0
\label{eq-gr2}\end{eqnarray}
where the integral ratios
\begin{eqnarray}
\tilde I_0&:=&{32\over3}{\int\psi\varphi\xi\dxi+3\int\varphi^2\theta^3\dxi\over\int{\beta\over\beta_c}\psi\xi^2\dxi}	\\
\tilde I_1&:=&{\int\omtilde^2\varphi\theta^3\xi\dxi\over\int{\beta\over\beta_c}\psi\xi^2\dxi}
\end{eqnarray}
are given uniquely by the dimensionless structure $(\xi,\theta,\varphi,\psi,\omtilde)$.
Notice that $\tilde I_0$ corresponds to the critical ratio $\beta_c/\sigma$ in the non-rotating case \citep{haemmerle2020c},
while $\tilde I_1$ gives the correction from rotation.
Equation~(\ref{eq-gr2}) has a unique positive root:
\begin{eqnarray}
\beta_{c,\rm crit}=\sqrt{{\kb T_c\over\mu\mh c^2}\tilde I_0+\left(2f^2\gtilde^{1/2}(T_c)\,\tilde I_1\right)^2}-2f^2\gtilde^{1/2}(T_c)\,\tilde I_1
\label{eq-betacrit}\end{eqnarray}
From equation~(\ref{eq-K}), the corresponding mass-scale (\ref{eq-mscale}) is given directly by $\beta_{c,\rm crit}$ and the chemical composition $\mu$:
\begin{equation}
4\pi\left({K_{\rm crit}\over\pi G}\right)^{3/2}={4\over\pi^{1/2}G^{3/2}}\left({3\over\asb}\right)^{1/2}\left({\kb\over\mu\mh}\right)^2\beta_{c,\rm crit}^{-2}
\label{eq-Kcrit}\end{equation}
Thus, the critical masses for the star and its core
\begin{eqnarray}
\Mcrit&=&4\pi\left({K_{\rm crit}\over\pi G}\right)^{3/2}\phisurf		\label{eq-Mcrit}\\
\Mcorecrit&=&4\pi\left({K_{\rm crit}\over\pi G}\right)^{3/2}\phicore	\label{eq-Mcorecrit}
\end{eqnarray}
are uniquely determined by the dimensionless structure $(\xi,\theta,\varphi,\psi,\omtilde)$ and the choice of $(f,T_c,\mu)$.

In the case $f=0$, when the stability limit is set by gas only, equation (\ref{eq-betacrit}) leads to $\beta_c/\sigma=\tilde I_0$.
In the opposite limit, when the stability is dominated by the effect of rotation,
\begin{eqnarray}
\sqrt{{\kb T_c\over\mu\mh c^2}\tilde I_0}\ll2f^2\gtilde^{1/2}(T_c)\,\tilde I_1,
\label{eq-rot}\end{eqnarray}
a linear development of equation~(\ref{eq-betacrit}) gives
\begin{eqnarray}
\beta_{c,\rm crit}\simeq{1\over4f^2}\left({\gbar\over\sqrt{16\pi G}}\right)^{1/2}\left({\asb\over3}\right)^{-1/4}{\kb\over\mu\mh c^2}{\tilde I_0\over\tilde I_1}
\label{eq-betarot}\end{eqnarray}
Notice that expression~(\ref{eq-betarot}) of $\beta_{c,\rm crit}$ is independent of $T_c$,
which implies that for large $f$ the critical mass-scale (\ref{eq-Kcrit}) is independent of $T_c$ and $\mu$:
\begin{eqnarray}
4\pi\left({K_{\rm crit}\over\pi G}\right)^{3/2}\simeq f^4{256\over\gbar}{c^4\over G}{\tilde{I}_1^2\over\tilde{I}_0^2}
\label{eq-mscalecrit}\end{eqnarray}
In other words, in the limit of large $f$, the critical mass of a given hylotrope is set uniquely by $f$.
This fraction appears to the power 4 in equation~(\ref{eq-mscalecrit}), which shows the strong dependence of the critical masses on the accreted angular momentum.
Moreover, it implies that for large $f$ the profile of the spin parameter (equations \ref{eq-spin}) at the limit of stability is independent of $f$,
and is given uniquely by the dimensionless structure:
\begin{eqnarray}
a_r={1\over4}\sqrt{\tilde{I}_0\over\tilde{I}_1}\,{\int_0^\varphi\Omtilde'\xi'^2\dif\varphi'\over\varphi^2}
\label{eq-spinuni}\end{eqnarray}
In particular, we have at the surface
\begin{eqnarray}
a_R={1\over 7}\sqrt{\tilde{I}_0\over\tilde{I}_1}\ \varphi_{\rm surf}^{-1/4}
\label{eq-087}\end{eqnarray}
We emphasise however that this universality is broken when gas plays a significant role in the determination of the stability limit.

Finally, we notice that the critical condition~(\ref{eq-betacrit}) can be expressed as a function of $E_\Omega/\vert W\vert$ (equation~\ref{eq-TWadim}) instead of $f$:
\begin{eqnarray}
\beta_{c,\rm crit}=\sqrt{{\kb T_c\over\mu\mh c^2}\tilde I_0+\left(2\tilde I_2{E_\Omega\over\vert W\vert}\right)^2}-2\tilde I_2{E_\Omega\over\vert W\vert}
\label{eq-betaTW}\end{eqnarray}
with
\begin{eqnarray}
\tilde I_2:={4\over3}{\int\varphi\theta^3\xi\dxi\over\int{\beta\over\beta_c}\psi\xi^2\dxi}
={\int\psi\xi^2\dxi\over\int{\beta\over\beta_c}\psi\xi^2\dxi}
\end{eqnarray}
The integral at the numerator has been transformed by an integration by parts, using equation~(\ref{eq-hydroadim}).

\section{Results}
\label{sec-res}

\subsection{Rotation profiles}
\label{sec-profiles}

\begin{figure}\begin{center}
\includegraphics[width=.5\textwidth]{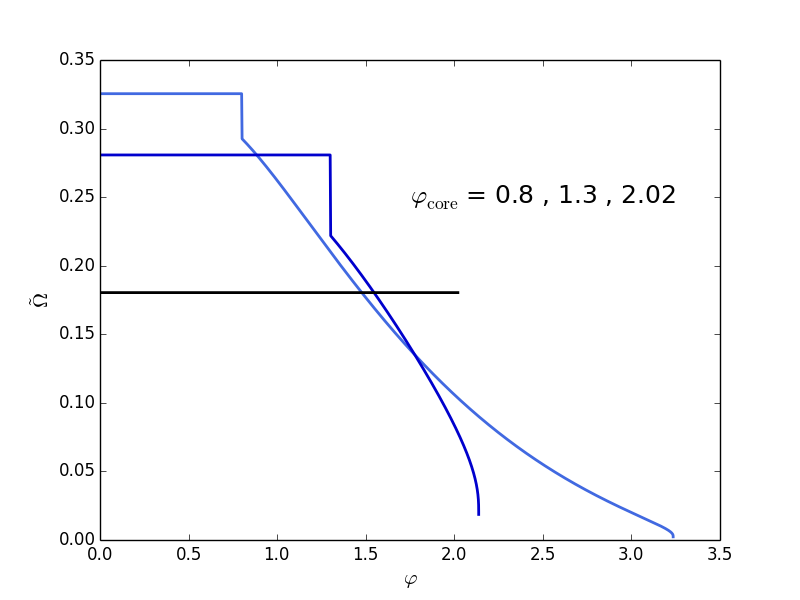}
\caption{Dimensionless rotation profiles of equation~(\ref{eq-Omtilde}), for a set of hylotropes with the indicated values of \phicore.}
\label{fig-omtilde}\end{center}\end{figure}

The dimensionless rotation profiles of equation~(\ref{eq-Omtilde}) are shown in figure~\ref{fig-omtilde} for a choice of hylotropes with the indicated values of \phicore.
The convective core, where solid rotation is imposed, appears clearly on each profile.
In particular, the case $\phicore=2.02$, which corresonds to the polytropic limit, is fully rotating as a solid body.
In contrast, strong differential rotation appears in the envelope of the other models,
and for $\phicore\lesssim1$ the angular velocity in the core exceeds by several orders of magnitude that of the surface,
consistently with the models accounting for full stellar evolution \citep{haemmerle2018b,haemmerle2019a}.
We see also that the instantaneous angular momentum transport in the core introduces a discontinuity in the \Omtilde-profiles at the interface with the envelope.
Physically, such infinite gradients shall be smoothed out by shears,
but here we can assume that such shears will lead only to a local redistribution of the angular momentum, without significant impact on the rotation profiles.

\begin{figure}\begin{center}
\includegraphics[width=.5\textwidth]{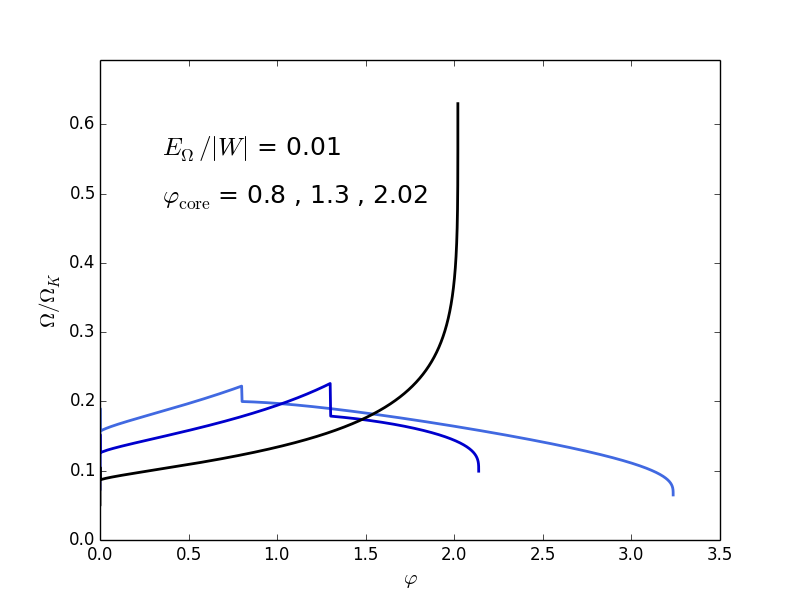}
\caption{Profiles of $\Omega/\OmK$ (equation~\ref{eq-omomkk}) for $E_\Omega/\vert W\vert=0.01$,
for a set of hylotropes with the indicated values of \phicore.}
\label{fig-omomk}\end{center}\end{figure}

From these dimensionless rotation profiles, the profiles of $\Omega/\OmK$ (equation~\ref{eq-omomk})
are given by the ratio $E_\Omega/\vert W\vert$ (equation~\ref{eq-omomkk}).
They are shown in figure~\ref{fig-omomk} for $E_\Omega/\vert W\vert=0.01$.
We see that the maximum values of $\Omega/\OmK$ are always reached at the surface of the convective core.
The rotation velocity remains typically 0.1 -- 0.2 \OmK, except in the very outer layers of the polytropic model ($\phicore=2.02$).
In this case, the \omgam-limit shall be exceeded at the surface, which shows that accretion onto fully convective SMSs requires $E_\Omega/\vert W\vert\ll0.01$.
On the other hand, the models with $\phicore\lesssim2$ keep a surface velocity $\sim0.1\OmK$ consistent with the \omgam-limit.
We see that, for a given constraint on the surface velocity, hylotropic structures allow for larger $E_\Omega/\vert W\vert$,
thanks to differential rotation in the envelope.
However, as already noticed in section~\ref{sec-omega}, the surface rotation of rapidly accreting SMSs
depends on the depth of their convective envelope, which is not included in the hylotropic structures.
The constraint on $f$ that arises from this effect is discussed in section~\ref{sec-mfin}.

\subsection{Critical masses}
\label{sec-masse}

\begin{figure}\begin{center}
\includegraphics[width=.5\textwidth]{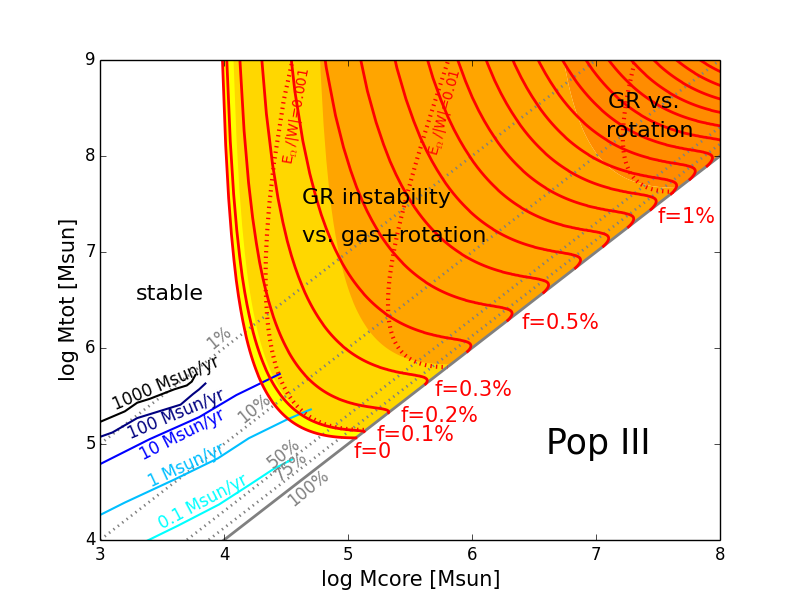}
\caption{Limits of stability for Pop III SMSs accreting angular momentum at a fraction $f$ of the Keplerian momentum.
The solid red lines show the hylotropic limits for the indicated $f$ (equation~\ref{eq-betacrit}),
with $T_c=1.8\times10^8$ K and $\mu=0.6$, relevant for Pop III SMSs.
The dotted red lines indicate the limit of stability for constant ratios 0.001 -- 0.01 -- 0.1 of rotational to gravitational energy (equation~\ref{eq-betaTW}).
The coloured areas reflect the relative contribution by gas and rotation in the determination of the stability limit~(\ref{eq-betacrit}):
from yellow to red, the rotation term $2f^2\gtilde^{1/2}(T_c)\,\tilde I_1$ represents respectively
$<0.1$, $0.1-1$, $1-10$ and $>10$ times the gas term $({\kb T_c\over\mu\mh c^2}\tilde I_0)^{1/2}$.
The Pop III \gva\ tracks of \cite{haemmerle2018a,haemmerle2019c} are shown for the indicated accretion rates.
The mass fraction of the convective core is indicated by the grey diagonals.}
\label{fig-mmcore}\end{center}\end{figure}

\begin{figure}\begin{center}
\includegraphics[width=.5\textwidth]{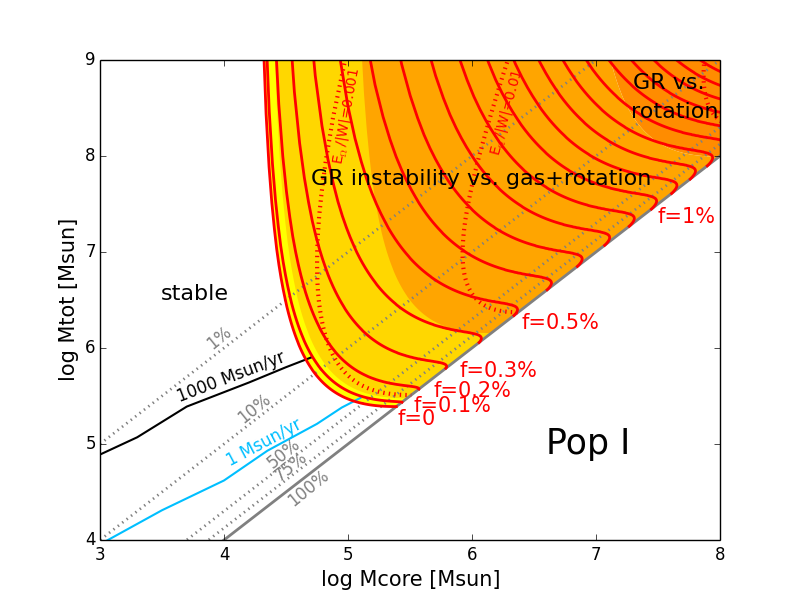}
\caption{Same as figure~\ref{fig-mmcore} for the Pop I case ($T_c=8.5\times10^7$ K).}
\label{fig-mmsol}\end{center}\end{figure}

The critical masses obtained with equations~(\ref{eq-Mcrit}-\ref{eq-Mcorecrit}) for the full series of hylotropes
are shown in figure~\ref{fig-mmcore} for $T_c=1.8\times10^8$ K, $\mu=0.6$, and a series of values of the fraction $f$.
These values of $T_c$ and $\mu$ are relevant for Pop III SMSs \citep{hosokawa2013,umeda2016,woods2017,haemmerle2018a,haemmerle2019c}.
In figure~\ref{fig-mmsol}, the critical masses are shown for the lower $T_c=8.5\times10^7$ K relevant for Pop I SMSs \citep{haemmerle2019c}.
For these central temperatures, the central densities are typically a few g cm$^{-3}$ for $M\sim10^5-10^6$ \Ms,
down to $\sim0.01-0.1$ g cm$^{-3}$ for $M\sim10^8-10^9$ \Ms.

Like in the non-rotating case \citep{haemmerle2020c}, for every given $f$,
the maximum total mass consistent with GR stability is a decreasing function of the core mass-fraction.
We see that the effect of rotation on the stability of the star becomes significant as soon as $f\gtrsim0.1\%$.
For $f=0.2-0.3\%$ the mass-limit is shifted by a significant factor, and for $f=0.3-0.5\%$ by an order of magnitude.
In this last case, the GR instability requires masses $\gtrsim10^6$ \Ms.
For $f=1\%$, the GR instability cannot be reached if the mass does not exceed several $10^7$ \Ms,
and the star can be stable with masses $\sim10^9$ \Ms\ if the core does not exceed $10^6$ \Ms.
In the range $f=1.5-2\%$, the star remains stable up to masses $10^8-10^9$ \Ms\ for any core mass.
We notice the small knee of the mass-limits near the polytropic structures ($\Mcore\gtrsim75\%M$).
The stabilising effect of rotation is slightly decreased by the solid rotation of the core,
that removes angular momentum from the densest central regions that are dominant for the instability.

The relative contribution by gas and rotation in the determination of the stability limit is made visible by coloured areas.
These contributions are estimated by the two terms that appear in the critical condition~(\ref{eq-betacrit}):
$({\kb T_c\over\mu\mh c^2}\tilde I_0)^{1/2}$, that represents gas, and $2f^2\gtilde^{1/2}(T_c)\,\tilde I_1$, that represents rotation.
The four regions, from yellow to red, correspond to the cases where the rotation term represents $<0.1$, $0.1-1$, $1-10$ and $>10$ times the gas term.
We see that, for $0.1\%\lesssim f\lesssim1\%$, both gas and rotation play a role in the determination of the stability limit.
But for $f\gtrsim1\%$, rotation becomes the only significant stabilising effect (condition~\ref{eq-rot}).
In this case, as noticed in section~\ref{sec-gr}, the critical masses of the various hylotropes are uniquely given by $f$,
independently of the thermal properties (equation~\ref{eq-mscalecrit}).
As a consequence, the effect of a change in the central temperature, between figures~\ref{fig-mmcore} and \ref{fig-mmsol},
impacts the mass-limits only for $f\lesssim1\%$, while the limits for $f\gtrsim1\%$ remain unchanged.

The mass-limits obtained for constant $E_\Omega/\vert W\vert$ (equation~\ref{eq-betaTW})
are shown in figures~\ref{fig-mmcore} and \ref{fig-mmsol} as red dotted lines, for $E_\Omega/\vert W\vert=0.001-0.01-0.1$.
Since the ratio $E_\Omega/\vert W\vert$ scales with $f^2$ (equation~\ref{eq-TWadim}),
the range of one order of magnitude in $f\sim0.1-1\%$ corresponds to a range of two orders of magnitude in $E_\Omega/\vert W\vert\sim0.001-0.1$.
For $\Mcore>10\%M$, the ratio $E_\Omega/\vert W\vert=0.01$ corresponds to $f\sim0.3-0.4\%$.
The fastest rotators shown in these figures ($f\sim1\%$) have $E_\Omega/\vert W\vert\sim0.1$,
which corresponds to typical velocities $\sim0.3\OmK$ (equation~\ref{eq-omomkk}).
We see that, for a given value of $E_\Omega/\vert W\vert$, hylotropic structures with small $\Mcore/M$
remain stable up to larger masses than polytropes ($\Mcore=M$).
More precisely, for a given relative contribution $E_\Omega/\vert W\vert$,
rotation has a stronger stabilising effect on hylotropes with $\Mcore<M$ than on polytropes.
It reflects the fact that rotational energy is more centralised when the core is smaller (figure~\ref{fig-omomk}),
and thus has a stronger impact on the dense regions that are relevent for the GR instability.

\subsection{Spin parameter}
\label{sec-spin}

The profile of the spin parameter is given by equation~(\ref{eq-spin}).
For each hylotrope, it is fully determined by the fraction $f$ and the mass-scale.
As notice in section~(\ref{sec-omega}), $f$ and the mass-scale appear in equation~(\ref{eq-spin}) as a global factor only,
so that a change in these quantities affects only the scale of the spin parameter, while the shape of the profile is unique for each hylotrope.
Here, we are interested in the profiles of the spin parameter at the limit of stability,
where the mass-scale is fixed by $f$, $T_c$ and $\mu$ (equations~\ref{eq-betacrit}-\ref{eq-Kcrit}).
Since the critical mass-scale is not the same for the various hylotropes,
the profiles of the spin parameter at the critical limit are rescaled by a different factor when $f$, $T_c$ and $\mu$ are changed.

\begin{figure}\begin{center}
\includegraphics[width=.5\textwidth]{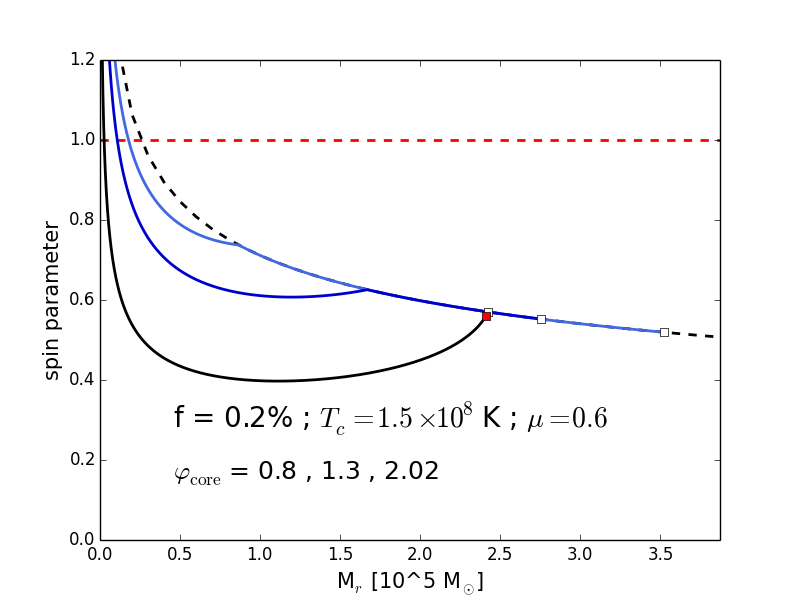}
\caption{Internal profile of the dimensionless spin parameter at the limit of stability for $f=0.2\%$, $T_c=1.5\times10^8$ K and $\mu=0.6$,
as a function of the mass-coordinate and for a choice of hylotropes with indicated \phicore.
The surface of each hylotrope is indicated by a white square.
The black dashed curve shows the angular momentum accretion law (\ref{eq-spinaccr}).
The red dashed curve and the red square indicate the limits for black hole formation, according to the criterions discussed in section~\ref{sec-spin2}.}
\label{fig-spin02}\end{center}\end{figure}

\begin{figure}\begin{center}
\includegraphics[width=.5\textwidth]{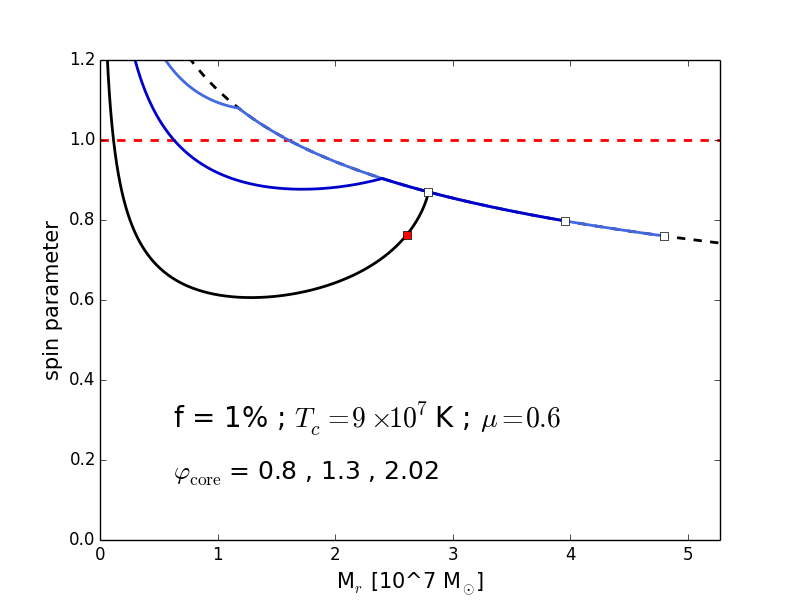}
\caption{Same as figure\ref{fig-spin02} for $f=1\%$ and $T_c=9\times10^7$ K.}
\label{fig-spin1}\end{center}\end{figure}

The critical profiles obtained for $f=0.2\%$, $T_c=1.5\times10^8$~K and $\mu=0.6$ are shown in figure~\ref{fig-spin02} for three hylotropes in the series.
The case $f=1\%$, $T_c=9\times10^7$ K and $\mu=0.6$ is shown in figure~\ref{fig-spin1}.
The profiles are compared with that of the angular momentum accretion law (\ref{eq-jmr}), expressed in equation~(\ref{eq-spinaccr}).
By construction, the profiles match each other in the radiative envelopes, and are impacted by the stellar structure only in the convective core.
By definition, the spin parameter diverges at the centre, but we see that for all the profiles it decreases below unity at a small fraction of the total mass.
The profile set by the accretion law, and kept in the envelope, is monotonously decreasing.
Thus, if the spin parameter goes below unity in the deep regions of the envelope, it will necessarily remain so in the whole envelope.
Only in the convective core, where solid rotation redistributes the angular momentum,
the spin parameter can increase outwards, in agreement with the polytropic models \citep{baumgarte1999b}.
However, it occurs only for hylotropes with $\phicore\gtrsim1$, i.e. for a core mass-fraction $\gtrsim40\%$.

We notice that in these two cases, $f=0.2\%$ and 1\%, the stability limit is set predominantly by gas and by rotation, respectively
(figures~\ref{fig-mmcore}-\ref{fig-mmsol}).
Condition~(\ref{eq-rot}) is already satisfied for $f=1\%$,
so that the critical profiles shown in figure~\ref{fig-spin1} correspond actually to the 'universal' profiles of equation~(\ref{eq-spinuni}),
independent of $f$ and given uniquely by the dimensionless structure.
As a consequence, these profiles represent an upper limit for each hylotrope,
and the increase of the critical masses for $f>1\%$ translates only into a rescaling along the $x$-axis.
In particular, in the case of the polytrope ($\phicore=2.02$),
we refind the profile of \cite{baumgarte1999b}, with the 'universal' value 0.87 at the surface, given by equation~(\ref{eq-087}).

\section{Discussion}
\label{sec-dis}

\subsection{Final mass as a function of the accretion history}
\label{sec-mfin}

Figures~\ref{fig-mmcore} and \ref{fig-mmsol} show that the maximum masses consistent with GR stability
depend sensitively on the fraction $f$ of accreted angular momentum, already in the range 0.1 -- 1 \%.
The actual value of $f$ at which a SMS can accrete is set by the \omgam-limit,
that imposes an upper limit for its surface velocity.
For stars with masses $\gtrsim10^5$ \Ms, the surface velocity cannot exceed $\sim10\%$ of the Keplerian velocity \citep{haemmerle2018b,haemmerle2019a}.
An extrapolation of the models suggests that the limit goes down to $\sim3\%$ for $\gtrsim10^7$~\Ms, and to $\sim1\%$ for $\gtrsim10^9$~\Ms.
As noticed in section~\ref{sec-omega}, for a given fraction $f$ of accreted angular momentum, the surface velocities of rapidly accreting SMSs
depend on the depth of their convective envelopes, which are not included in the hylotropic models.
Models accounting for full stellar evolution show that this envelope covers more than 2/3 of the total stellar radius for rates $\lesssim1$ \Mpy,
which implies $f\lesssim0.2-0.3\%$ \citep{haemmerle2019a}.
On the other hand, the depth of the envelope is significantly reduced for rates $\gtrsim100$ \Mpy,
and covers a small fraction of the total radius.
As long as it remains the case,
the SMS can accrete angular momentum at $f\sim1\%$ up to $\sim10^9$ \Ms\ without facing the \omgam-limit.

Rapidly accreting SMSs are invoked in two different versions of direct collapse: atomically cooled haloes and galaxy mergers.
The accretion rates in atomically cooled haloes are expected to be $0.1-10$ \Mpy\ most probably $\lesssim1$ \Mpy\ \citep{latif2013e,chon2018,patrick2020},
which imposes $f\lesssim0.2-0.3\%$.
On the other hand, galaxy mergers allows for rates $\gtrsim100$ \Mpy\ \citep{mayer2010,mayer2015,mayer2019},
for which $f\simeq1\%$ remains consistent with the \omgam-limit up to large masses.
An other difference between these two scenarios is the chemical composition, which is primordial in atomically cooled haloes,
while solar metallicities are expected in the case of galaxy mergers.
Black hole formation at the collapse of a Pop I SMS requires a mass $\gtrsim10^6$ \Ms\ \citep{montero2012}.
In the non-rotating case, such final masses could be reached only for rates $\gtrsim1000$ \Mpy\ \citep{haemmerle2020c,haemmerle2021a}.

The evolutionary tracks of the \gva\ models accounting for full stellar evolution \citep{haemmerle2018a,haemmerle2019c},
computed under accretion at 0.1 -- 1000 \Mpy\ up to masses $10^5-10^6$~\Ms, are plotted in figures~\ref{fig-mmcore} and \ref{fig-mmsol}.
These models are built on the assumption of hydrostatic equilibrium, and thus are insensitive to dynamical instabilities.
Their stability with respect to GR has been addressed in \cite{haemmerle2021a} for the non-rotating case, by the direct use of equation~(\ref{eq-chandra}).
As shown in \cite{haemmerle2020c}, the hylotropic limit reproduces these final masses with a high precision
for accretion rates $\geq10$ \Mpy\ and remains a good approximation for lower rates.
We see that none of these tracks reach the stability limits for $f\gtrsim0.2\%$.
An extrapolation of the tracks suggests that, for conditions of atomically cooled haloes (Pop~III, $\dm\sim0.1-1$ \Mpy, $f\sim0.2-0.3\%$),
rotation allows to increase the final mass by a factor $\sim2$ only.
It implies that the mass of SMSs forming in this scenario remain always $<10^6$ \Ms, more probably $\lesssim5\times10^5$ \Ms.
On the other hand, for the conditions of metal-rich galaxy mergers (Pop~I, $\dm\gtrsim100$ \Mpy, $f\sim1\%$),
rotation could allow SMSs to increase their final mass by several orders of magnitude, leading to an upper mass-limit of $10^8-10^9$~\Ms.
Overall, like in the non-rotating case, the final masses of SMSs range in distinct intervals in the two versions of direct collapse:
$10^5-10^6$ \Ms\ for atomically cooled haloes and $10^6-10^9$ \Ms\ in galaxy mergers.
We notice also that the minimum rate required for final masses $\gtrsim10^6$ \Ms\ is slightly reduced by rotation,
since for 100 \Mpy\ already, we can expect $f\sim1\%$.

Interestingly, an extrapolation of the tracks at $\dm\gtrsim100~$\Mpy\ in figures~\ref{fig-mmcore} and \ref{fig-mmsol}
suggests that stars accreting at these rates are mostly convective when they reach the GR instability for $f\simeq1\%$.
The same is true for $\dm\lesssim0.1$ \Mpy\ and $f=0.2\%$.
If the star becomes fully convective before the instability, the instantaneous angular momentum transport affects the surface velocity,
and the values of $f$ consistent with the \omgam-limit are decreased below $<0.1$.
Thus, either the accreted angular momentum decreases, so that the star becomes rapidly GR unstable, or the surface reaches break-up.
In this case, accretion must stop and the star evolves along the mass-shedding limit,
in a similar pathway as monolithic models \citep{fowler1966,bisnovatyi1967,baumgarte1999b}.
In this case, only second-order post-Newtonian corrections have been found to trigger the collapse \citep{bisnovatyi1967}.
We emphasise however that, in the accretion scenario considered here, these second-order corrections remain always negligible,
and we checked numerically that they do not play a role in the determination of the limits in figures~\ref{fig-mmcore} and \ref{fig-mmsol}.
Notice also that the age of a star accreting at 100 -- 1000 \Mpy\ when it reaches $10^8-10^9$ \Ms\ is $\sim10^6$ years,
at which H-burning might be achieved \citep{umeda2016,woods2021}.
In this case, pair-instability might be reached before the GR instability \citep{woods2020}.

\subsection{Angular momentum accretion law and rotation profiles}
\label{sec-rotation}

The definition of the rotation profiles in section~\ref{sec-omega}
relies on the angular momentum accretion law~(\ref{eq-jac}) and the mass-radius relation~(\ref{eq-rac}).
This relation is verified by stellar evolution models for masses $\lesssim10^5$~\Ms\ \citep{hosokawa2013,haemmerle2018a},
but for larger masses the models depart towards smaller radii.
In the absence of stellar evolution models accounting for accretion up to masses $\geq10^6$~\Ms,
the mass-radius relation is not known for the most massive objects considered in the present work.
According to equation~(\ref{eq-jfmr}), smaller radii would imply a slower rotation.
Notice however that the radius appears in this equation through its square-root,
so that a decrease by an order of magnitude in radius would be equivalent to a decrease by a factor of a few only in $f$.
Moreover, the mass-radius relation of the zero-age main-sequence (ZAMS) imposes a lower limit to the radius.
Like the Hayashi line, the ZAMS of SMSs follows a nearly constant effective temperature \citep{woods2020},
leading to the same power-law $R\propto M^{1/2}$ as equation~(\ref{eq-rac}), rescaled down by 2 -- 3 orders of magnitude in radius.
Such a shift would be equivalent to a decrease of $f$ by one order of magnitude.
Since departures from relation~(\ref{eq-rac}) might appear only for $M\gtrsim10^6$~\Ms,
we can estimate that this effect would translate in the $(\Mcore,M)$ diagram of figures~\ref{fig-mmcore} and \ref{fig-mmsol}
into a shift of the mass-limits for $f\gtrsim0.5\%$, down to that for $f=0.4\%$.
We notice however that more recent models of very-massive stars
show a shift in the ZAMS towards lower effective temperatures and larger radii \citep{grafener2021}.
This effect might inhibit a possible contraction of SMSs in the mass-range $\gtrsim10^6$ \Ms,
allowing for accretion of relatively high angular momentum.

In addition to the mass-radius relation, the definition of the rotation profiles in section~\ref{sec-omega}
relies on the assumption of local angular momentum conservation in the envelope, which leads to strong differential rotation.
Stellar models \citep{haemmerle2018b} have shown shear diffusion and meridional circulation to be inefficient in rapidly accreting SMSs,
because of the short timescales and slow rotations.
On the other hand, magnetic fields (e.g. a Taylor-Spruit dynamo), that are neglected in the present work, could inhibit significantly differential rotation.
In the extreme case of solid rotation, the constraint from the \omgam-limit at the surface translates into $f<0.1\%$,
and the effect of rotation on the stability becomes negligible.
In other words, for given surface velocities,
solid rotation would impose a low angular momentum in the the deep regions of the star that are decisive for the GR instability,
and could bring back the stability limit to the non-rotating case.

We notice also that the ratio $\Omega/\OmK$ remains $<0.7$ for all the rotation profiles considered ($f\lesssim2\%$),
except in the very outer layers of the nearly-polytropic models ($\Mcore/M>0.95$), which play negligible role in the GR instability.
No significant flattening is expected for such slow rotations \citep{ekstroem2008a}, which justifies the use of the spherical hylotropic structures.
Moreover, in the deep regions that are relevent for the GR instability, the ratio $\Omega^2/\OmKK$ remains typically 0.1 -- 0.2.
Thus, we can estimate that post-Newtonian rotation terms in the pulsation equation~(\ref{eq-postnewtonomega}),
that are neglected in the present method, should remain small compared to the other terms.

\subsection{Spin parameter and black hole formation}
\label{sec-spin2}

The spin parameter (\ref{eq-spindef}) is key for the direct formation of a supermassive black hole during the collapse of a SMS.
Without angular momentum transport, the profile $a_r(M_r)$ at the onset of the instability is conserved during the collapse.
For monolithic SMSs, evolving along the mass-shedding limit and fully stabilised by rotation (condition~\ref{eq-rot}),
analytical models in the post-Newtonian limit feature a universal profile of the spin parameter,
with in particular a unique value 0.87 at the surface \citep{baumgarte1999b}.
Except in the central $\sim10\%$ of the mass, the spin remains always $a_r<1$, the maximum value allowed for the formation of a Kerr black hole,
which suggests that the whole stellar mass can be accreted by the black hole in a dynamical time.
Notice that, since monolithic models are rotating at the Keplerian limit, the conditions for a Newtonian treatment of rotation are not satisfied,
and the full GR numerical simulations of \citep{baumgarte1999b} lead to larger values for the spin parameter (0.97).
Only the slow rotations imposed by accretion along the \omgam-limit guarantee the validity of the Newtonian treatment of rotation against GR instability.

The hydrodynamical simulations of the collapse of monolithic SMSs have shown that
the outer $\sim10\%$ of the stellar mass remains in orbit outside the horizon \citep{shibata2002,shapiro2002,liu2007,uchida2017,sun2017,sun2018}.
The resulting non-axisymmetric structure is found to trigger gravitational wave emission and ultra-long gamma-ray bursts,
which are currently the main observational signatures of the existence of SMSs proposed in the literature for future detection.
\cite{shapiro2002} showed that the mass of the orbiting structure in the simulations can be derived analytically
by comparing the specific angular momentum of each layer $M_r$
with that of the innermost stable circular orbit (ISCO) of a black hole with mass $M_r$ and spin $a_r$:
if in a layer the angular momentum exceeds the ISCO value, the layer remains in orbit outside the horizon.

The critical profiles of the spin parameter shown in figures~\ref{fig-spin02} and \ref{fig-spin1}
correspond to typical conditions of atomically cooled haloes and galaxy-mergers, respectively.
For conditions of galaxy mergers, we refind the profile of \cite{baumgarte1999b} for the polytrope ($\phicore=2.02$),
which reflects the fact that condition~(\ref{eq-rot}) is already satisfied for $f=1\%$.
In contrast, for the case of atomically cooled haloes ($f=0.2\%$, figure~\ref{fig-spin02}),
gas plays the dominant role in setting the stability limit (figure~\ref{fig-mmcore}), and the spin parameter departs from the 'universal' profiles,
as already noticed by \cite{shibata2016a} and \cite{butler2018} for monolithic models.
By comparing the angular momentum profiles with the ISCO values of the various layers (equation~17 of \citealt{shapiro2002}),
we can estimate in which cases an outer envelope rotates fast enough to remain in circular orbit after black hole formation.
The limit of this envelope is shown by a red square in the profiles, and we see that it occurs only for nearly polytropic models.
For $f=1\%$, it requires a core mass-fraction $\gtrsim80\%$ and in the polytropic limit we refind that the outer 10\% of the stellar mass can remain in a circular orbit.
For $f=0.2\%$ the core mass-fraction must exceed $\gtrsim99\%$ and even in the polytropic limit
it is only the outer $\sim1\%$ of the mass that can stand in orbit.

These results indicate that the angular momentum barrier never prevents the direct formation of a supermassive black hole at the collape of a SMS.
They suggest that the star must be mostly convective for a significant mass-fraction to remain outside the horizon at the formation of the black hole.
Moreover, the mass-fraction of such outer structures are found much larger under the conditions of galaxy mergers than under those of atomically cooled haloes.
Thus, the conditions of galaxy mergers appear as more promising than those of atomically cooled haloes
for detectable gravitational wave emission and ultra-long gamma-ray bursts during the collapse.

\section{Summary and conclusions}
\label{sec-out}

We have extended our previous works on the GR instability in rapidly accreting SMSs \citep{haemmerle2020c,haemmerle2021a},
based on the relativistic pulsation equation of \cite{chandrasekhar1964},
by including rotation according to the method of \cite{fowler1966}.
On the basis of hylotropic models relevent for rapidly accreting SMSs \citep{begelman2010,haemmerle2019c,haemmerle2020c},
we defined rotation profiles by assuming local angular momentum conservation in radiative regions, which allows for strong differential rotation.
The angular momentum advected by each layer at accretion is assumed to represent a fraction $f$ of the Keplerian angular momentum.

The impact of rotation on the GR instability is captured by a positive term in the pulsation equation,
which translates into a stabilising effect, like gas pressure, against the destabilising GR effects.
The stabilising effect of rotation becomes significant as soon as $f\gtrsim0.1\%$.
For $f\sim0.2\%$, the maximum masses consistent with GR stability are increased by a significant factor compared to the non-rotating case.
For $f\sim1\%$, it is increased by several orders of magnitude, and the GR instability cannot be reached if the mass does not exceed $10^7-10^8$ \Ms.

The relevent values of $f$ depend on the channel of direct collapse:
while $f\sim0.2\%$ is expected in atomically cooled haloes, because of the deep convective envelope of SMSs accreting at rates $\lesssim1$ \Mpy,
values $f\sim1\%$ appear consistent in conditions of galaxy mergers, thanks to the larger rates.
It indicates that rotation allows for SMSs forming in atomically cooled haloes to increase their final masses by a factor of a few only.
In contrast, SMSs forming in galaxy mergers could increase their masses by several orders of magnitude compared to the non-rotating case,
possibly up to $\sim10^9$ \Ms.
It follows that the final masses of rapidly accreting SMSs range in distinct intervals in the different versions of direct collapse:
$10^5-10^6$~\Ms\ for primordial, atomically cooled haloes; $10^6-10^9$~\Ms\ for metal-rich galaxy mergers.

The rotational properties of rapidly accreting SMSs at the onset of GR instability are not universal, in contrast to the monolithic, non-accreting case.
The monolithic profiles are found in the limit of fully convective SMSs with $f\gtrsim1$ \%,
which illustrates the fact that, already for such low $f$, rotation plays the dominant role in the stability of the star.
We estimate that the angular momentum barrier is inefficient to prevent the direct formation of a supermassive black hole during the collapse.
It is only in the limit of fully convective SMSs with $f\gtrsim1\%$ that a significant fraction of the stellar mass ($\sim10\%$) can remain in orbit outside the horizon.
Thus, the conditions of galaxy mergers appear as more promising than those of atomically cooled haloes
regarding the possibility for detectable gravitational wave emission and ultra-long gamma-ray bursts during the collapse.

\appendix

\section{Newtonian pulsation analysis with slow rotation}
\label{app-euler}

We consider a Newtonian star in slow rotation, so that spherical symmetry can be assumed.
Each layer $r$ of the star rotates with a given angular velocity $\Omega(r)$, i.e. an average specific angular momentum:
\begin{equation}
j=\langle r^2\sin^2\theta\ \Omega\rangle={\int r^2\sin^2\theta\ \Omega\,\sin\theta\,\dth\dphi\over\int\sin\theta\,\dth\dphi}={2\over3}r^2\Omega
\label{eq-j}\end{equation}
The average is done by integration with respect to the solid angle $\sin\theta\,\dth\dphi$ over the whole sphere $r$,
and the additional $\sin^2\theta$ factor that appears in the numerator comes from the rotation radius $r\sin\theta$ squared.
In this case, the average of the radial, centrifugal acceleration is:
\begin{equation}
\langle r\sin^2\theta\ \Omega^2\rangle={\int r\sin^2\theta\ \Omega^2\,\sin\theta\,\dth\dphi\over\int\sin\theta\,\dth\dphi}={2\over3}r\Omega^2
={3\over2}{j^2\over r^3}
\label{eq-centri}\end{equation}
The two $\sin\theta$ factors in the averaged quantity come from the rotational radius $r\sin\theta$
and the projection of the centrifugal force on the radial direction.

We assume an equilibrium state with centrifugal acceleration~(\ref{eq-centri})
and denote the corresponding quantities with a subscript $_0$:
\begin{equation}
0=-{GM_r\over r_0^2}-4\pi r_0^2{\dP_0\over\dMr}+{3\over2}{j^2\over r_0^3}
\label{eq-equilibrium}\end{equation}
In equation~(\ref{eq-equilibrium}), we adopted the Lagrangian mass as coordinate, so that the subsript can be omitted for $M_r$.
We consider small periodic perturbations to this equilibrium state,
with local angular momentum conservation, which justifies the omission of the subscript for $j$.
The dynamical quantities satisfy the Euler equation with centrifugal acceleration (\ref{eq-centri}):
\begin{equation}
\ddot r=-{GM_r\over r^2}-4\pi r^2{\dP\over\dMr}+{3\over2}{j^2\over r^3}
\label{eq-euler}\end{equation}
The amplitude of the periodic perturbations is expressed as an unknown function $\epsilon(M_r)\ll1$, and the pulsation frequency $\omega$ is assumed to be constant:
\begin{eqnarray}
r&=&r_0(1+\epsilon e^{i\omega t})			\label{eq-dr}
\end{eqnarray}
From expression~(\ref{eq-dr}) of $r$, the density is given by the equation of continuity (\ref{eq-continu}), after linearisation:
\begin{eqnarray}
\rho={1\over4\pi r^2{\dr\over\dMr}}=\rho_0\left({1-{1\over r_0^2}{\dif\left(r_0^3\epsilon\right)\over\dr_0}e^{i\omega t}}\right)
\label{eq-drho}\end{eqnarray}
The pertubations are assumed to be adiabatic, so that the pressure changes are related to the density changes by the first adiabatic exponent:
\begin{eqnarray}
{P-P_0\over P_0}=\Gamma_1{\rho-\rho_0\over\rho_0}	\ll1
\end{eqnarray}
With equation~(\ref{eq-drho}), it gives
\begin{eqnarray}
P=P_0\left(1-{\Gamma_1\over r_0^2}{\dif\left(r_0^3\epsilon\right)\over\dr_0}e^{i\omega t}\right)
\label{eq-dP}\end{eqnarray}
For the perturbation~(\ref{eq-dr}-\ref{eq-drho}-\ref{eq-dP}), a linearisation of equation~(\ref{eq-euler}) gives:
\begin{eqnarray}
-\omega^2r_0\epsilon e^{i\omega t}=
-{GM_r\over r_0^2}(1-2\epsilon e^{i\omega t})-4\pi r_0^2{\dP_0\over\dMr}(1+2\epsilon e^{i\omega t})	\\
+4\pi r_0^2{\dif\over\dMr}\left({\Gamma_1P_0\over r_0^2}{\dif\left(r_0^3\epsilon\right)\over\dr_0}\right)e^{i\omega t}
+{3\over2}{j^2\over r_0^3}(1-3\epsilon e^{i\omega t})
\end{eqnarray}
The subscript $_0$ can now be omitted, since all quantities refer to equilibrium except $\epsilon$ and $\omega$.
Equation (\ref{eq-equilibrium}) allows to cancel the constant terms and the $e^{i\omega t}$ factors:
\begin{eqnarray}
\omega^2r\epsilon=16\pi r^2{\dP\over\dMr}\epsilon+{3\over2}{j^2\over r^3}\epsilon
-4\pi r^2{\dif\over\dMr}\left({\Gamma_1P\over r^2}{\dif\left(r^3\epsilon\right)\over\dr}\right)
\end{eqnarray}
If we use the continuity equation (\ref{eq-continu}) to replace $\dif/\dMr$ by $\dif/\dr$,
and if we multiply all the terms by $r^2$, the equation takes the following form:
\begin{eqnarray}
\omega^2r^3\epsilon={r^2\over\rho}\left(\left({4\over r^3}{\dP\over\dr}+{3\over2}{\rho j^2\over r^6}\right)r^3\epsilon
-{\dif\over\dr}\left({\Gamma_1P\over r^2}{\dif\left(r^3\epsilon\right)\over\dr}\right)\right)
\label{eq-propre}\end{eqnarray}
Equation~(\ref{eq-propre}) is an eigenvalue problem for a self-adjoint linear operator.
A sufficient condition for dynamical instability can be found by projecting this equation over the function $r^3\epsilon$
according to the corresponding scalar product:
\begin{eqnarray}
\omega^2\int\rho r^4\epsilon^2\dr=4\int r^3{\dP\over\dr}\epsilon^2\dr+{3\over2}\int\rho j^2\epsilon^2\dr	\\
-\int r^3\epsilon{\dif\over\dr}\left({\Gamma_1P\over r^2}{\dif\left(r^3\epsilon\right)\over\dr}\right)\dr
\end{eqnarray}
The first and last integrals in the right-hand side can be transformed by integrations by parts:
\begin{eqnarray}
\omega^2\int\rho r^4\epsilon^2\dr=-4\int P{\dif(r^3\epsilon^2)\over\dr}\dr+{3\over2}\int\rho j^2\epsilon^2\dr	\\
+\int{\Gamma_1P\over r^2}\left({\dif\left(r^3\epsilon\right)\over\dr}\right)^2\dr
\end{eqnarray}
If there is a function $\epsilon$ that satisfies this equation for $\omega^2<0$,
it implies that real exponentials $e^{\vert\omega\vert t}$ are solutions of the Euler equation~(\ref{eq-euler}), and the star is unstable.
The simplest case is a homologous perturbation, where $\delta r\propto r$,
i.e. $\epsilon^2=\cst$ can be extracted out of the integrals and cancelled in both sides:
\begin{eqnarray}
\omega^2\int\rho r^4\dr&=&9\int\left(\Gamma_1-{4\over3}\right)Pr^2\dr+{3\over2}\int\rho j^2\dr
\end{eqnarray}
Using equation~(\ref{eq-j}) to replace $j$ by $\Omega$, and the equation of continuity~(\ref{eq-continu}) to replace \dr\ by \dMr, we obtain
\begin{eqnarray}
\omega^2I&=&6\int\left(\Gamma_1-{4\over3}\right)P\dif V+{4\over9}\int\Omega^2r^2\dMr	\\
&=&6\int\left(\Gamma_1-{4\over3}\right)P\dif V+{4\over9}\int{\Omega^2\over\OmKK}{GM_r\over r}\dMr
\end{eqnarray}
where $\dif V=\dMr/\rho=4\pi r^2\dr$ is the volume element, \OmK\ is the Keplerian velocity (equation~\ref{eq-kepler}) and
\begin{eqnarray}
I=\int\langle r^2\sin^2\theta\rangle\dMr={2\over3}\int r^2\dMr
\end{eqnarray}
is the total moment of inertia.

In the Eddington limit, $\Gamma_1=4/3+\beta/6+\mathcal{O}(\beta^2)$, and we are left with
\begin{eqnarray}
\omega^2I&=&\int\beta P\dif V+{4\over9}\int{\Omega^2\over\OmKK}{GM_r\over r}\dMr	\label{eq-puls1}\\
&=&\int\pgaz\dif V+{4\over9}\int{\Omega^2\over\OmKK}{GM_r\over r}\dMr				\label{eq-puls2}
\end{eqnarray}
Finally, we notice that these two integrals can be rewritten in terms of internal energy of gas and rotational energy:
\begin{eqnarray}
U\gaz=\int{3\over2}\pgaz\dif V
\end{eqnarray}
\begin{eqnarray}
E_\Omega=\int{1\over2}\langle r^2\sin^2\theta\ \Omega^2\rangle\dMr&=&{1\over3}\int r^2\Omega^2\dMr	\\
&=&{1\over3}\int{\Omega^2\over\OmKK}{GM_r\over r}\dMr
\label{eq-Eom}\end{eqnarray}
so that equation~(\ref{eq-puls2}) reads:
\begin{eqnarray}
\omega^2I={2\over3}U_\gaz+{4\over3}E_\Omega
\end{eqnarray}

\begin{acknowledgements}
LH has received funding from the European Research Council (ERC) under the European Union's Horizon 2020 research and innovation programme
(grant agreement No 833925, project STAREX).
\end{acknowledgements}

\bibliographystyle{aa}
\bibliography{bib}

\end{document}